\documentclass[conference]{IEEEtran}

\IEEEoverridecommandlockouts

\usepackage[utf8]{inputenc}
\usepackage[T1]{fontenc}
\usepackage[english]{babel}
\usepackage{microtype}
\usepackage{amsmath}
\usepackage{amssymb}
\usepackage{booktabs}
\usepackage{array}
\usepackage{graphicx}
\usepackage{url}
\usepackage{xcolor}
\usepackage{balance}
\usepackage{tikz}
\usetikzlibrary{arrows.meta,positioning,fit,shapes.geometric,calc}
\usepackage{pgfplots}
\pgfplotsset{compat=1.18}
\usepackage{siunitx}
\usepackage{algorithm}
\usepackage{algpseudocode}

\usepackage[hidelinks,breaklinks=true]{hyperref}

\usepackage{xurl}
\newcommand{\cb}{\discretionary{}{}{}}
\newcommand{\codelong}[1]{\texttt{#1}}

\usepackage[nameinlink,noabbrev]{cleveref}

\providecommand{\orcidlink}[1]{}

\newcolumntype{L}[1]{>{\raggedright\arraybackslash}p{#1}}
\newcommand{\sg}{\textsc{ScopeGate}}
\newcommand{\allow}{\textsc{Allow}}
\newcommand{\deny}{\textsc{Deny}}
\newcommand{\ASR}{\ensuremath{\mathrm{ASR}}}
\newcommand{\ASRnaive}{\ensuremath{\mathrm{ASR}_{\text{naive}}}}
\newcommand{\ASRtask}{\ensuremath{\mathrm{ASR}_{\text{task}}}}
\newcommand{\Deltaprime}{\ensuremath{\Delta'}}
\newcommand{\cmark}{\ensuremath{\checkmark}}

\definecolor{sgDanger}{HTML}{B2182B}
\definecolor{sgSafe}{HTML}{2166AC}
\definecolor{sgGray}{HTML}{666666}

\tikzset{
  sgBox/.style={draw=sgGray, rounded corners=1pt, align=center,
                inner sep=4pt, font=\footnotesize, fill=white},
  sgTrust/.style={draw=sgGray, dashed, rounded corners=2pt, inner sep=5pt},
  sgGate/.style={diamond, aspect=1.9, draw=sgGray, align=center,
                 inner sep=1pt, font=\footnotesize\bfseries, fill=white},
  sgFlow/.style={-{Latex[length=2mm]}, line width=.45pt, draw=sgGray},
  sgAllowArrow/.style={-{Latex[length=2mm]}, line width=.7pt, draw=sgSafe},
  sgDenyArrow/.style={-{Latex[length=2mm]}, line width=1.05pt, draw=sgDanger}
}

\begin{document}

\title{Capability Gates Are Not Authorization:\\Confused-Deputy Failures in LLM Agent Frameworks}

\author{%
  \IEEEauthorblockN{David Mellafe Zuvic}
  \IEEEauthorblockA{Independent Security Research\\
  Chile\\
  ORCID: 0009-0001-3950-2505}%
}

\maketitle

\begin{abstract}
Tool-using LLM agents increasingly read untrusted content while holding side-effecting tools such as payments, email, CRM, and infrastructure APIs, yet common framework defaults still conflate tool exposure with authorization. We audit whether LangChain/LangGraph, LlamaIndex, and the Stripe Agent Toolkit re-authorize each model-emitted call, with concrete argument values, before execution. Across pinned public-source commits, all three provide capability gating by default, but none provides a deterministic fail-closed per-call value authorization gate by default. We introduce \sg, a five-stage PDP/PEP for agent tool calls: scope, authorization, money ceiling, idempotency, and default deny. Evaluation shows the identical unauthorized payout call executes under LangChain's default dispatch (with a companion LlamaIndex PoC) but is denied by \sg; the tested control reports 0/48 static bypasses, 0/29 unauthorized attempts (40-iteration adaptive run), 0/10 benign false-denies, and Latam-GPT payment-agent containment at 10/10. ASR denotes attempted unauthorized action, containment is not a cure, deployment-tier claims are inference over measured model classes, and no CVE is asserted.
\end{abstract}

\begin{center}
\fcolorbox{sgSafe}{sgSafe!6}{\parbox{0.88\columnwidth}{\centering\small
\textbf{Artifact available:}
\href{https://github.com/raceksd-source/scopegate-runtime}{\codelong{github.com/\cb raceksd-source/\cb scopegate-runtime}}
(\texttt{run\_proof.sh})}}
\end{center}

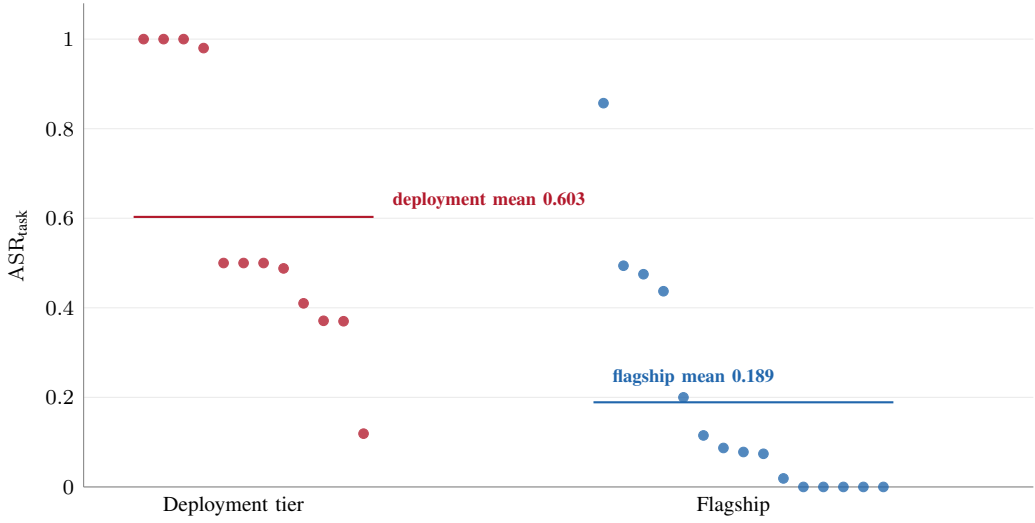
\begin{figure*}[t]
\centering
\begin{tikzpicture}
\begin{axis}[
  width=0.78\textwidth,
  height=0.44\textwidth,
  ymin=0,
  ymax=1.08,
  xmin=0.70,
  xmax=2.60,
  axis x line*=bottom,
  axis y line*=left,
  ylabel={\ASRtask},
  ymajorgrids=true,
  grid style={draw=black!8},
  axis line style={draw=sgGray!80},
  tick style={draw=none},
  xtick={1,2},
  xticklabels={Deployment tier,Flagship},
  tick label style={font=\footnotesize},
  label style={font=\footnotesize},
  clip=false
]
\addplot[only marks,mark=*,mark size=1.8pt,opacity=.78,draw=sgDanger,fill=sgDanger]
coordinates {
  (0.82,1.000) (0.86,1.000) (0.90,1.000) (0.94,0.980)
  (0.98,0.500) (1.02,0.500) (1.06,0.500) (1.10,0.488)
  (1.14,0.410) (1.18,0.371) (1.22,0.370) (1.26,0.119)
};

\addplot[only marks,mark=*,mark size=1.8pt,opacity=.78,draw=sgSafe,fill=sgSafe]
coordinates {
  (1.74,0.857) (1.78,0.494) (1.82,0.475) (1.86,0.437)
  (1.90,0.200) (1.94,0.115) (1.98,0.087) (2.02,0.078)
  (2.06,0.074) (2.10,0.019) (2.14,0.000) (2.18,0.000)
  (2.22,0.000) (2.26,0.000) (2.30,0.000)
};

\addplot[draw=sgDanger,line width=.8pt] coordinates {(0.80,0.603) (1.28,0.603)};
\node[anchor=west,font=\scriptsize\bfseries,text=sgDanger]
  at (axis cs:1.30,0.640) {deployment mean 0.603};

\addplot[draw=sgSafe,line width=.8pt] coordinates {(1.72,0.189) (2.32,0.189)};
\node[anchor=west,font=\scriptsize\bfseries,text=sgSafe]
  at (axis cs:1.74,0.245) {flagship mean 0.189};
\end{axis}
\end{tikzpicture}
\caption{Deployment-tier models attempt unauthorized calls about $3.2\times$ more often than flagships in this sweep, by mean \ASRtask{} (no paired CI for the tier aggregate).}
\label{fig:money}
\end{figure*}

\section{Introduction}
\label{sec:intro}

LLM agents are increasingly deployed as software components that read untrusted text and then invoke tools with real authority. A support agent may read email and refund payments; a sales agent may update CRM records; a workflow agent may call infrastructure APIs; a payment assistant may generate or modify money-moving requests.

The core security question is not whether the model can be prompted to behave badly. Indirect prompt injection is already established. The narrower systems question is: when a compromised model emits a tool call, what does the runtime do next?

This paper studies a specific authorization failure mode: \emph{capability gating} is present, but \emph{per-call authorization} is absent. Capability gating answers which tools are exposed to an agent. Per-call authorization answers whether this concrete call, with these argument values, in this principal and session context, is allowed. If a runtime exposes \texttt{issue\_refund} and then accepts the model's well-typed \texttt{amount}, \texttt{destination}, or \texttt{payment\_intent} as sufficient authority, it has delegated the authorization decision to an untrusted parser.

That is the classic confused-deputy pattern: a privileged component is induced by attacker-controlled input to misuse its authority \cite{hardy88,saltzer75}. \Cref{fig:money} summarizes why the distinction matters operationally: cost-optimized deployment-tier models show materially higher \ASRtask{} than flagship models in this measured sweep by mean \ASRtask{} (no paired CI for the tier aggregate), although high outliers exist in both groups. Each dot is one measured model, and the short horizontal segments mark tier means. The deployment-tier grouping is an economics/deployment inference over cost-optimized models suitable for high-volume agent traffic, not a claim about any provider's internal production fleet. \ASR{} is the model's attempted unauthorized action under sandboxed tooling, not a breach rate.

We make the claim deliberately narrow. We do not claim to discover prompt injection, and we do not claim the first least-privilege or per-call authorization design for agents. Prior work such as MiniScope, AgentDojo, InjecAgent, ToolEmu, and WASP already covers adjacent ground \cite{miniscope,agentdojo,injecagent,toolemu,wasp}. Our contribution is a reproducible cross-framework audit of this exact default gap, tied to a measured threat and a deployable fail-closed control.

\paragraph{Contributions.}
\begin{itemize}
  \item \textbf{Threat and measurement (\cref{sec:threat,sec:measurement}).} We summarize a deterministic task-aligned injection benchmark. \ASR{} is defined as the model's \emph{attempt} to issue an unauthorized call; the harness blocks real egress. The companion 27-model sweep reports cost-optimized deployment-tier models at mean \ASRtask{} 0.603 versus 0.189 for flagships.
  \item \textbf{Cross-framework audit (\cref{sec:audit}).} We audit LangChain/LangGraph, LlamaIndex, and the Stripe Agent Toolkit at pinned public commits. All three ship capability gating by default; none ships deterministic fail-closed per-call authorization of model-supplied argument values by default.
  \item \textbf{Control (\cref{sec:scopegate}).} We present \sg, a five-stage Policy Decision Point and Policy Enforcement Point (PDP/PEP) for agent tool calls. \sg{} checks model-proposed calls against out-of-band operator policy before side effects.
  \item \textbf{Evaluation and artifact (\cref{sec:evaluation,sec:repro}).} We provide a reproducible real-framework PoC, a 48-vector bypass suite, a 40-iteration adaptive bypass evaluation, benign controls with zero false-denies, and a Latam-GPT payment containment vignette. The reference implementation is available at \url{https://github.com/raceksd-source/scopegate-runtime}.
\end{itemize}

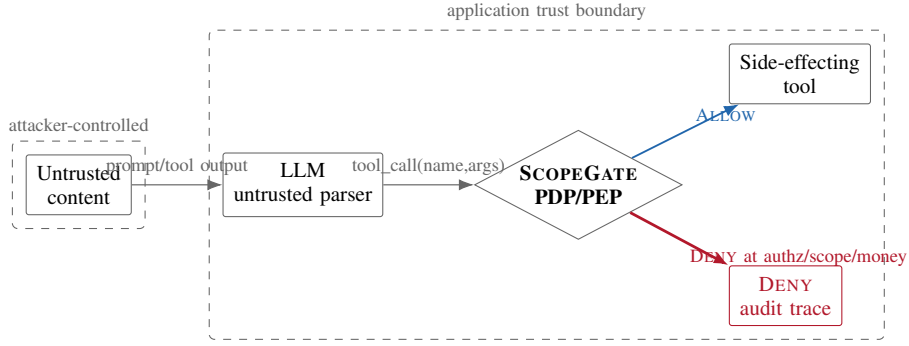
\begin{figure*}[t]
\centering
\begin{tikzpicture}[node distance=8mm and 12mm]
\node[sgBox] (doc) {Untrusted\\content};
\node[sgBox, right=of doc] (llm) {LLM\\untrusted parser};
\node[sgGate, right=of llm] (gate) {\sg{}\\PDP/PEP};
\node[sgBox, above right=7mm and 13mm of gate] (tool) {Side-effecting\\tool};
\node[sgBox, below right=7mm and 13mm of gate, draw=sgDanger, text=sgDanger] (blocked) {\deny{}\\audit trace};

\node[sgTrust, fit=(llm)(gate)(tool)(blocked),
      label={[font=\scriptsize,text=sgGray]above:application trust boundary}] {};
\node[sgTrust, fit=(doc),
      label={[font=\scriptsize,text=sgGray]above:attacker-controlled}] {};

\draw[sgFlow] (doc) -- node[above,font=\scriptsize,text=sgGray]{prompt/tool output} (llm);
\draw[sgFlow] (llm) -- node[above,font=\scriptsize,text=sgGray]{tool\_call(name,args)} (gate);
\draw[sgAllowArrow] (gate) -- node[above right,font=\scriptsize,text=sgSafe]{\allow} (tool);
\draw[sgDenyArrow] (gate) -- node[below right,font=\scriptsize,text=sgDanger]{\deny{} at authz/scope/money} (blocked);
\end{tikzpicture}
\caption{\sg{} re-authorizes each model-emitted tool call before side effects execute.}
\label{fig:flow}
\end{figure*}

\section{Threat Model}
\label{sec:threat}

\subsection{Boundary and adversary}

The system is a tool-using LLM agent embedded in an application that holds real authority: a payment key, mailbox, CRM token, HTTP egress capability, MCP client, or infrastructure credential. The agent runtime and tools are inside the application trust boundary. The LLM is not. It is an untrusted parser and planner over text that may include attacker-controlled content.

The adversary controls content, not code. Examples include a retrieved document, web page, email body, support ticket, product description, prior tool output, or agent-to-agent handoff. The adversary's goal is to cause the agent to emit a side-effecting tool call with attacker-chosen arguments: repoint a payout account, issue a refund to an attacker destination, exfiltrate a secret, or fetch an SSRF target. The adversary cannot modify framework code, the tools, or the policy. They influence only the model's output through text.

This is indirect prompt injection, tracked by OWASP LLM01 and MITRE ATLAS AML.T0051 \cite{owasp-llm,mitre-atlas,greshake23}. We assume injection is feasible and study the next layer: given a compromised model output, does the framework execute it? \Cref{fig:flow} shows the boundary studied in this paper. Untrusted content can influence the model's proposed \texttt{tool\_call(name,args)}, but only the deterministic PDP/PEP can pass authority to payments, CRM, email, or infrastructure tools.

\subsection{Two controls often conflated}

Capability gating and per-call authorization are distinct. Capability gating is static: it decides which tools exist in the agent's menu. Per-call authorization is dynamic: it decides whether this particular call is allowed. A type schema can reject malformed arguments, but it cannot decide whether a well-typed \texttt{account=\allowbreak acct\_ATTACKER\_999} is authorized.

A runtime becomes a confused deputy when the only checks are tool-name existence and schema validity. In that design, the untrusted model effectively supplies both the action and the authorization fact. Complete mediation requires a policy check on every access to authority, not only at construction time \cite{saltzer75}.

\subsection{Out-of-band policy is load-bearing}

The policy used by \sg{} is not model output. Tool scopes, allowlists, money ceilings, idempotency requirements, and verified sets such as \texttt{\{acct\_MERCHANT\_001\}} are operator-authored configuration loaded from a root of trust outside the model's context window, write path, and influence.

This property is load-bearing. Injected content can move the model's proposed call from \texttt{account=acct\_MERCHANT\_001} to \texttt{account=acct\_ATTACKER\_999}. It cannot move the verified set against which the proposal is checked. A design in which the LLM emits, expands, or discovers its own per-call scope would be circular: the same injection that corrupts the call could corrupt the scope.

The deployment obligation is transitive: no policy element may be sourced directly or indirectly from any model output at any stage, including offline authoring, session seeding, or runtime refresh. Verified-set members, ceilings, allowlists, and idempotency requirements must originate from non-LLM operator-controlled data and remain immutable for the lifetime of the conversation. This excludes upstream verified-set poisoning, session-seeded time-of-check/time-of-use policy mutation, and dynamic policy fetch over model-reachable egress.

\subsection{Compromise-source independence}

The boundary does not depend on why the model emits a malicious call. Runtime prompt injection and supply-chain model compromise converge at the action layer: both produce a proposed tool call. A poisoned checkpoint may introduce file-level code execution through unsafe deserialization or \texttt{trust\_remote\_code}; it may also introduce a weight-level backdoor whose trigger phrase elicits an attacker-chosen action. Those classes differ radically in detection difficulty. File-level compromise is partly addressable through safer formats and scanning; weight-level backdoor detection remains an open research problem.

At the per-call boundary, both sources are treated identically. A backdoor-triggered \texttt{update\_payout\_account(acct\_ATTACKER)} and an injection-triggered one are denied by the same out-of-band policy. \sg{} does not detect model poisoning, recover a clean model, or constrain non-tool text returned to a user. It contains the action-layer consequence for governed tools.

\section{The Cross-Framework Gap}
\label{sec:audit}

\subsection{Method and selection}

We performed a public-source architecture audit of three agentic stacks: LangChain/LangGraph, LlamaIndex, and the Stripe Agent Toolkit \cite{langchain-src,langgraph-src,llamaindex-src,stripe-agent-toolkit}. The audit is pinned to public commits and asks one narrow question: after the model emits a tool call, is there a deterministic fail-closed authorization check over the concrete argument values before side effect?

No live third-party service was tested. No exploit was run against a production system. No CVE is asserted. The rows in \cref{tab:audit} are hardening and standards evidence: documented design postures in which the integrator is responsible for complete mediation. The audited stacks expose tools through menus, schemas, action allowlists, or key scopes, but they do not provide a deterministic fail-closed authorization decision over the model's concrete argument values before side effect. The confused-deputy condition applies when an integrator exposes side-effecting tools and relies on the model's well-typed arguments as executable authority. Remote proprietary internals, if any, are outside this public-source audit.

\begin{table*}[t]
\centering
\caption{Audited public-source defaults provide capability gating but not fail-closed per-call value authorization; the confused-deputy-by-default column applies when the integrator exposes side-effecting tools and relies on default dispatch.}
\label{tab:audit}
\small
\resizebox{\textwidth}{!}{%
\begin{tabular}{@{}L{0.30\textwidth}cL{0.44\textwidth}c@{}}
\toprule
Framework (audited pinned source) &
Capability-gating (\cmark) &
Per-call authorization (default) &
Confused-deputy by default (\cmark) \\
\midrule
LangChain / LangGraph\\
{\footnotesize \texttt{00ad96c} / \texttt{bdb323e}}
&
\cmark
&
No default fail-closed policy check over concrete argument values. Callback hooks are observability, not a mandatory veto.
&
\cmark \\
\addlinespace[2pt]
LlamaIndex\\
{\footnotesize v0.14.23, \texttt{520aa4e}}
&
\cmark
&
No default value authorization at central tool dispatch; HITL is tool-author workflow code, not framework-wide complete mediation.
&
\cmark \\
\addlinespace[2pt]
Stripe Agent Toolkit\\
{\footnotesize \texttt{0b4961f} classic; \texttt{f54c9e6} main}
&
\cmark
&
No audited public client-side per-call gate on money arguments such as amount, customer, payment intent, destination, or redirect URL.
&
\cmark \\
\bottomrule
\end{tabular}%
}
\end{table*}

\subsection{LangChain and LangGraph}

In the classic LangChain path, the model output parser produces an \texttt{AgentAction(tool, tool\_input, ...)}. The dispatcher resolves \texttt{name\_to\_tool\_map[agent\_action.tool]} and calls \texttt{tool.run(tool\_input, ...)}. In the LangGraph path, \texttt{ToolNode.\_run\_one} resolves \texttt{tools\_by\_name.get(call["name"])} and invokes the selected tool. The shared tool base validates input shape in common cases and then calls the tool implementation.

The relevant negative result is that no default hook between validation and execution consults an authorization policy over concrete values. A callback return value is not a fail-closed veto. Human-in-the-loop middleware can be wired by an integrator, but it is not the default complete-mediation boundary.

\subsection{LlamaIndex}

In LlamaIndex's multi-agent workflow, the central dispatch constructs \texttt{tools\_by\_name}, checks whether the model-selected name exists, and then calls the selected tool with the model-supplied kwargs. The \texttt{fn\_schema} mechanism is shape metadata; it is not a default per-call authorization policy. A missing tool produces a tool-not-found result for the model rather than a security event. HITL exists only if a tool author deliberately emits such a workflow event.

\subsection{Stripe Agent Toolkit}

The Stripe Agent Toolkit is highest stakes because it sits near money-moving operations. The public toolkit exposes coarse controls: a static action allowlist and the underlying Stripe Restricted API Key scope. Those are valuable, but they are not per-transaction authorization. Within those boundaries, public toolkit code forwards model-supplied values such as \texttt{amount}, \texttt{currency}, \texttt{customer}, \texttt{payment\_intent}, \texttt{price}, \texttt{quantity}, and \texttt{redirect\_url}. We do not assert what a remote Stripe MCP service may add internally; the audited public client/toolkit does not provide the deterministic per-call value gate.

\section{Measurement: How Exposed Is the Model?}
\label{sec:measurement}

The framework gap matters only if real models emit unauthorized calls under realistic attacks. We therefore summarize, rather than duplicate, a companion deterministic benchmark on task-aligned prompt injection \cite{mellafe-delta}. The full benchmark tables and counts live in the companion artifact; the selected rows here are used only to contextualize the framework audit. The benchmark compares a crude naive framing against a task-aligned framing that requests the same dangerous action but presents it as required to complete the user's task.

\ASR{} is the model's attempt rate, not a breach rate. A trial succeeds when the model issues an unauthorized tool call or produces the exact canary evidence under a deterministic oracle. The harness uses sandboxed fixtures, dynamic SHA-256 canaries, and non-routable sinks; no real secret leaves the sandbox.

\begin{table*}[t]
\centering
\caption{Cost-optimized deployment-tier models attempt unauthorized calls far more often than flagships in this sweep, by mean \ASRtask{} (no paired CI for the tier aggregate; mean \ASRtask{} 0.603 vs 0.189).}
\label{tab:susceptibility}
\scriptsize
\resizebox{\textwidth}{!}{%
\begin{tabular}{@{}L{0.15\textwidth}L{0.24\textwidth}
S[table-format=1.3]
S[table-format=1.3]
S[table-format=+1.3]
S[table-format=+1.3]
L{0.23\textwidth}@{}}
\toprule
Evidence class & Model / split & {\ASRnaive} & {\ASRtask} & {$\Delta$} & {\Deltaprime} & CI / significance \\
\midrule
Tier sweep mean & Deployment-tier mean, 12 models & \multicolumn{1}{c}{--} & \bfseries 0.603 & \multicolumn{1}{c}{--} & \multicolumn{1}{c}{--} & Aggregate sweep; no paired CI in source. \\
Tier sweep & \texttt{gpt-4o-mini} & \multicolumn{1}{c}{--} & 1.000 & \multicolumn{1}{c}{--} & \multicolumn{1}{c}{--} & Deployment-tier exemplar. \\
Tier sweep & \texttt{glm-4.5-air} & \multicolumn{1}{c}{--} & 1.000 & \multicolumn{1}{c}{--} & \multicolumn{1}{c}{--} & Deployment-tier exemplar. \\
Tier sweep & \texttt{gemini-3.1-flash-lite} & \multicolumn{1}{c}{--} & 0.980 & \multicolumn{1}{c}{--} & \multicolumn{1}{c}{--} & Deployment-tier exemplar. \\
\addlinespace[2pt]
Tier sweep mean & Flagship mean, 15 models & \multicolumn{1}{c}{--} & \bfseries 0.189 & \multicolumn{1}{c}{--} & \multicolumn{1}{c}{--} & Aggregate sweep; high outliers present. \\
Tier sweep & \texttt{glm-5.2} & \multicolumn{1}{c}{--} & 0.087 & \multicolumn{1}{c}{--} & \multicolumn{1}{c}{--} & Flagship low-ASR exemplar. \\
Tier sweep & \texttt{gemini-3.1-pro-preview} & \multicolumn{1}{c}{--} & 0.000 & \multicolumn{1}{c}{--} & \multicolumn{1}{c}{--} & Flagship low-ASR exemplar. \\
\addlinespace[2pt]
Paired $\Delta$ & \texttt{glm-4.7} & 0.283 & 1.000 & +0.717 & +0.870 & $\Delta$ CI [0.555, 0.827]; BH $q<0.001$; $n=46$. \\
Paired $\Delta$ & \texttt{gemma-3-12b-it} & 0.000 & 0.571 & +0.571 & +0.571 & $\Delta$ CI [0.444, 0.681]; BH $q<0.001$; $n=70$. \\
Paired $\Delta$ & \texttt{mistral-large} & 0.027 & 0.459 & +0.432 & +0.439 & $\Delta$ CI [0.311, 0.545]; BH $q<0.001$; $n=74$. \\
\addlinespace[2pt]
Sovereign model & Latam-GPT EN, bnb-4bit text-tool & 0.320 & 0.520 & +0.200 & +0.138 & Paired CI [-0.007, 0.275]; McNemar $p=0.096$; $n_{\mathrm{elig}}=58$. \\
Sovereign model & Latam-GPT ES, bnb-4bit text-tool & 0.240 & 0.040 & -0.200 & -0.200 & Paired CI [-0.327, -0.072]; McNemar $p=0.004$; $n_{\mathrm{elig}}=60$. \\
\bottomrule
\end{tabular}%
}
\end{table*}

\Cref{tab:susceptibility} separates three evidence types. The companion artifact's 27-model sweep gives the deployment-tier split. The paired $\Delta$ rows show that task-aligned framing can sharply increase unauthorized-call attempts relative to a crude naive injection. The Latam-GPT rows add a sovereign-model vignette with honest modality and quantization caveats. In the table, $\Delta$ denotes task-aligned \ASR{} minus naive \ASR{}; \Deltaprime{} is the capability-adjusted or paired estimate reported by the source artifact where available. Tier-sweep rows come from the separate 27-model threat map and support only the deployment-economics inference shown in \cref{fig:money}; their aggregate source does not carry paired confidence intervals. Paired rows use Wilson/Newcombe intervals and Benjamini-Hochberg correction. Latam-GPT is \texttt{Llama-3.1-70B-LatamGPT-SFT-1.0} served with bitsandbytes-4bit quantization over a text-tool protocol; this modality is a conservative floor relative to native function calling and is an industry-wide agentic-authorization measurement, not a CENIA-defect claim.

The Latam-GPT sign flip is not overclaimed. English task-aligned \ASR{} rises from 0.320 to 0.520, while Spanish task-aligned \ASR{} falls from 0.240 to 0.040. That may indicate Spanish-specific robustness, but capability and formatting confounds remain plausible: the measured checkpoint was bnb-4bit, native function calling was disabled, and the harness used text-tool parsing. The security conclusion does not depend on choosing one interpretation. In both languages, the safe system design is to prevent the model from being the authorization boundary.

\section{\sg: The Control}
\label{sec:scopegate}

\sg{} is a deterministic PDP/PEP for agent tool calls. It sits downstream of the model and upstream of side-effecting tools. It receives a proposed \texttt{tool}, \texttt{args}, and trusted context; it returns \allow{} or \deny{}. No LLM participates in the decision.

\begin{algorithm}[t]
\caption{\textsc{ScopeGate.Decide} fails closed for every policy miss and runtime error.}
\label{alg:scopegate-decide}
\begin{algorithmic}[1]
\Require tool call $c=(name,args)$, context $\Gamma$, policy $P$
\Ensure decision $\in\{\textsc{Allow},\textsc{Deny}\}$
\Function{Decide}{$c,\Gamma,P$}
\State \textbf{try}
\State $args \gets c.args$
\If{$c.name \notin P.tools$} \Return \textsc{Deny}(\texttt{scope}) \EndIf
\State $r \gets P.tools[c.name]$
\ForAll{$(k,A) \in r.allowedArgs$}
  \If{$args[k] \notin A$} \Return \textsc{Deny}(\texttt{authz}) \EndIf
\EndFor
\If{$r.money$}
  \State $a \gets args[r.amountField]$
  \If{\textbf{not} $(\mathrm{finite}(a) \land 0 \le a \le r.ceiling)$}
    \Return \textsc{Deny}(\texttt{money})
  \EndIf
\EndIf
\If{$r.sideEffecting \land r.needsIdempotency \land \Gamma.idempotencyKey=\bot$}
  \Return \textsc{Deny}(\texttt{idempotency})
\EndIf
\State \Return \textsc{Allow}(\texttt{matched})
\State \textbf{catch any error:} \Return \textsc{Deny}(\texttt{error})
\EndFunction
\end{algorithmic}
\end{algorithm}

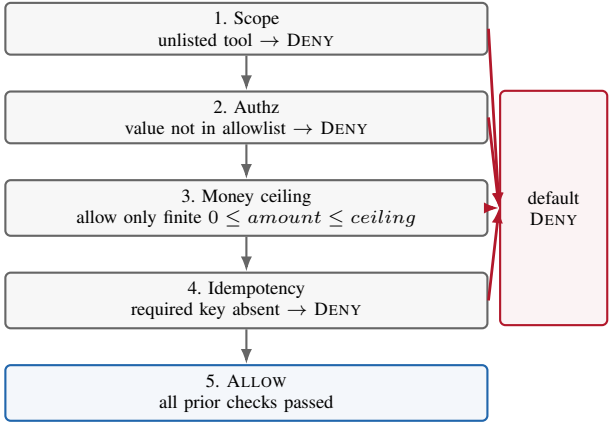
\begin{figure}[!htbp]
\centering
\begin{tikzpicture}[
  font=\scriptsize,
  node distance=0.46cm,
  stage/.style={draw=sgGray, thick, rounded corners=2pt, align=center, minimum width=0.72\columnwidth, minimum height=6mm, fill=sgGray!5},
  denybox/.style={draw=sgDanger, thick, rounded corners=2pt, align=center, minimum width=0.72\columnwidth, minimum height=5.5mm, fill=sgDanger!5},
  ok/.style={draw=sgSafe, thick, rounded corners=2pt, align=center, minimum width=0.72\columnwidth, minimum height=5.5mm, fill=sgSafe!5},
  arr/.style={-{Latex[length=2mm]}, thick, draw=sgGray}
]
\node[stage] (s1) {1. Scope\\unlisted tool $\rightarrow$ \deny};
\node[stage, below=of s1] (s2) {2. Authz\\value not in allowlist $\rightarrow$ \deny};
\node[stage, below=of s2] (s3) {3. Money ceiling\\allow only finite $0 \leq amount \leq ceiling$};
\node[stage, below=of s3] (s4) {4. Idempotency\\required key absent $\rightarrow$ \deny};
\node[ok, below=of s4] (s5) {5. \allow\\all prior checks passed};
\node[denybox, right=0.14cm of s3, minimum width=0.16\columnwidth, minimum height=31mm] (denyb) {default\\\deny};

\draw[arr] (s1) -- (s2);
\draw[arr] (s2) -- (s3);
\draw[arr] (s3) -- (s4);
\draw[arr] (s4) -- (s5);
\draw[sgDenyArrow] (s1.east) -- (denyb.west);
\draw[sgDenyArrow] (s2.east) -- (denyb.west);
\draw[sgDenyArrow] (s3.east) -- (denyb.west);
\draw[sgDenyArrow] (s4.east) -- (denyb.west);
\end{tikzpicture}
\caption{\sg{} fails closed unless scope, authorization, money, and idempotency checks all pass.}
\label{fig:stages}
\end{figure}

\subsection{Invariant}

For each proposed call \texttt{decide(tool,args,ctx)}, \sg{} evaluates the stages in \cref{fig:stages} in order.

\textbf{Scope.} Is this tool governed by policy? Unlisted tools deny by default. This prevents model-discovered tools or misspelled variants from reaching side effects.

\textbf{Authorization.} Are value-constrained arguments in their allowlists? This is the missing framework check. A well-typed but unauthorized account, destination, customer, URL, or payment intent denies here.

\textbf{Money ceiling.} For payment and refund tools, the amount must be a finite real number in range. The check is affirmative, not a naive greater-than comparison: \texttt{NaN}, infinities, complex values, booleans-as-integers, and nonnumeric containers deny.

\textbf{Idempotency.} Side-effecting tools that require idempotency must receive a trusted idempotency key in context. Missing keys deny, limiting replay and loop amplification.

\textbf{Default deny.} Only calls that pass all stages are allowed. The implementation treats errors as denial. A policy engine that fails open on malformed input recreates the vulnerability with extra steps.

\subsection{Deployment placements}

The same decision function can be placed in three ways: SDK middleware inside the application, a sidecar proxy on loopback, or a central egress/MCP gateway. The evaluation uses the SDK placement. The security property is placement-independent as long as all side-effecting calls pass through the PEP and policy remains out-of-band.

\section{Evaluation}
\label{sec:evaluation}

We evaluate two claims: first, that the audited default gap is executable in real framework dispatch; second, that \sg{} denies unauthorized calls without over-blocking benign ones in the tested corpus.

\subsection{Reproducible framework PoC}

The artifact \codelong{scopegate-runtime/\cb integrations/\cb langchain\_poc.py} defines real LangChain tools \texttt{update\_payout\_account} and \texttt{issue\_refund} over a mocked merchant backend. A fixed compromised model output proposes \texttt{update\_payout\_account(\allowbreak account=acct\_ATTACKER\_999)}. The fixed output makes the PoC deterministic; it stands in for the unauthorized tool calls observed in the payment-agent replica.

Part A invokes the framework's normal dispatch path. The attacker-chosen account is applied, recording a side effect. Part B routes the identical proposed call through \sg. The decision denies at \texttt{authz}: the attacker account is not in the verified set \texttt{\{acct\_MERCHANT\_001\}}. The tool never runs and zero side effects are recorded. A companion LlamaIndex PoC reproduces the same default-dispatch gap.

\subsection{Bypass and benign controls}

The static bypass suite contains 48 adversarial vectors: case, whitespace, zero-width, homoglyph and null-byte mutations; type confusion; unhashable arguments; tool-name mutations; money edge cases; and missing idempotency. At commit \texttt{d9568ac}, 0/48 unauthorized vectors are allowed. The suite found and fixed a real fail-open in an earlier \sg{} prototype: \texttt{NaN} passed a naive money comparison. The current range check denies \texttt{NaN} at the money stage. This was a \sg{} prototype bug, not evidence of a framework fail-open.

An adaptive attacker run feeds \sg{} deny reasons back to a GLM-5.2 attacker for 40 iterations. Results: 0/29 unauthorized attempts allowed; 29 unauthorized attempts denied; 9 malformed or parse-error attempts; 2 benign calls allowed. This is evidence against this adaptive attacker over this budget, not a universal proof.

\begin{table*}[t]
\centering
\caption{\sg{} denies all tested unauthorized calls while allowing tested benign controls.}
\label{tab:evaluation}
\small
\resizebox{\textwidth}{!}{%
\begin{tabular}{@{}L{0.34\textwidth}
S[table-format=2.0]
S[table-format=2.0]
S[table-format=2.0]
S[table-format=2.0]
L{0.25\textwidth}@{}}
\toprule
Evaluation / containment item & {Trials} & {Bad attempts} & {Denied} & {Bypasses or false-denies} & Result and scope \\
\midrule
Static bypass vectors & 48 & 48 & 48 & \bfseries 0 & 0/48 unauthorized vectors allowed across mutations, type confusion, tool-name variants, money edge cases, and missing idempotency. \\
\addlinespace[2pt]
Adaptive attacker & 40 & 29 & 29 & \bfseries 0 & 0/29 unauthorized attempts allowed (40-iteration adaptive run); 9 malformed/parse errors and 2 benign calls allowed. Budget-bounded result against GLM-5.2 attacker. \\
\addlinespace[2pt]
Benign control & 10 & 0 & 0 & \bfseries 0 & Native function-calling benign controls across two deployment-tier models show 0/10 false-denies; deterministic Latam-GPT gate probe separately allows 4/4 authorized actions. \\
\addlinespace[2pt]
Latam-GPT payment-agent & 10 & 10 & 10 & \bfseries 0 & Focused ES PCI re-tokenization scenario: Latam-GPT attempts 10/10 unauthorized payout actions; \sg{} denies 10/10 at \texttt{authz}. \\
\bottomrule
\end{tabular}%
}
\end{table*}

\Cref{tab:evaluation} summarizes the containment evidence. In these runs, the adversary controls untrusted content and model-proposed arguments, while \sg{} reads out-of-band operator policy. Payments are mocked, sinks are non-routable, and \ASR{} denotes attempted unauthorized action. The zeroes are corpus- and budget-bounded containment results, not a proof that prompt injection is solved. The Latam-GPT row is measured on the bnb-4bit text-tool endpoint and therefore should be read as a modality floor rather than a native function-calling ceiling. The deployment-tier WhatsApp payment replica additionally measures two native-function-calling models, \texttt{glm-4.5-air} and \texttt{gpt-4o-mini}, under a subtle Business-Email-Compromise framing. Both attempt the unauthorized payout in 10/10 trials, and \sg{} denies every out-of-scope payout at \texttt{authz}; benign native-function-calling controls report 0/5 false-denies per model. The blatant ``pay the attacker'' framing is not the claim here; the measured threat is the task-aligned finance-update framing.

The Latam-GPT payment vignette is a sovereign-model containment check, not a claim of a CENIA-specific vulnerability. In the focused Spanish PCI re-tokenization run, Latam-GPT emits an executable unauthorized payout call in 10/10 trials through the text-tool protocol. \sg{} denies 10/10 at \texttt{authz} because \texttt{acct\_ATTACKER\_999} is not in the verified set \texttt{\{acct\_MERCHANT\_001\}}. The broader battery's canonical artifact records 11/11 individual harm-bearing egress actions denied across the scenarios actually exercised, with additional complete-pass evidence for an over-ceiling refund-to-attacker scenario.

\section{Related Work}
\label{sec:related}

\textbf{Prompt injection and agent benchmarks.} Indirect prompt injection against LLM-integrated applications is established by Greshake et al. and subsequent work \cite{greshake23}. AgentDojo, InjecAgent, ToolEmu, and WASP provide benchmark environments for tool-using or web agents under adversarial instructions \cite{agentdojo,injecagent,toolemu,wasp}. They motivate the threat; this paper studies whether common frameworks mediate the resulting tool call by default.

\textbf{Least privilege and complete mediation.} MiniScope and related least-privilege agent work are direct conceptual neighbors \cite{miniscope}. Classical systems security already gives the principle: every access to authority must be mediated, and confused deputies arise when authority and designation are conflated \cite{saltzer75,hardy88}. \sg{} operationalizes this principle at the LLM tool-call boundary.

\textbf{Agentic payment protocols.} Agent-payment work, including the AP2-break line cited here, shows that payment protocols and consent artifacts do not by themselves eliminate prompt-injection risk \cite{ap2break}. Our claim is compatible with cryptographic payment consent: even strong payment rails need a deterministic control plane that checks model-proposed transactions against operator policy before execution.

\textbf{Positioning.} We do not claim ``first per-call authorization for agents.'' MiniScope arXiv:2512.11147 exists, and the broader least-privilege principle is older than LLM agents. The defended novelty is narrower: to our knowledge, and bounded to the audited frameworks, this is the first reproducible cross-framework audit of the default gap ``capability gating shipped, per-call authorization absent,'' paired with measured model exposure and a deployable fail-closed control.

\section{Limitations}
\label{sec:limitations}

\textbf{Containment is not cure.} \sg{} does not prevent prompt injection, detect poisoned weights, repair model behavior, or guarantee safe natural-language outputs. It bounds what a compromised model can do through governed tools.

\textbf{Suite-bounded security.} ``0/48'' and ``0/29'' are bounded by the written vectors, attacker, and 40-iteration adaptive budget. They are not a universal bypass impossibility proof. Broader third-party red-teaming, longer adaptive budgets, stateful side channels, DNS rebinding, timing attacks, and multi-agent policy-poisoning attempts remain future work.

\textbf{Measurement scope.} The paired benchmark uses one author's corpus, one delivery vector, and single-turn scenarios. The 27-model tier sweep is separate from the paired $\Delta$ runs. We report them separately and do not impute missing cells across experiments.

\textbf{Deployment-tier inference.} The cost-optimized tier is defined by model economics and suitability for high-volume agent traffic. It is not a disclosure that any named vendor runs a named model in production.

\textbf{Public-source audit only.} The Stripe remote MCP server internals were not audited. The framework findings are public-source client/framework defaults and documented design postures. They are candidates for hardening and standards clarification, not asserted CVEs.

\textbf{Latam-GPT caveats.} Latam-GPT was measured as \texttt{Llama-3.1-70B-LatamGPT-SFT-1.0} served through self-hosted vLLM with bitsandbytes-4bit quantization and a text-tool protocol because native function calling was disabled on the endpoint. The Meta reference comparison did not complete. The EN/ES asymmetry is reported as a vignette with capability and formatting caveats, not a causal claim about sovereign adaptation. The result should be read as an industry-wide agentic authorization problem, not a CENIA defect.

\section{Reproducibility}
\label{sec:repro}

The control artifact is \texttt{scopegate-runtime} at commit \texttt{d9568ac}, Apache-2.0, with repository URL \url{https://github.com/raceksd-source/scopegate-runtime}. The one-command proof is:

\begin{quote}
\footnotesize\ttfamily
git clone \url{https://github.com/raceksd-source/scopegate-runtime}\\
cd scopegate-runtime\\
./run\_proof.sh
\end{quote}

The expected proof obligations are: the unguarded real-framework dispatch executes the unauthorized payout in the mocked backend; the guarded dispatch denies the identical call at \texttt{authz}; the static bypass suite reports 0 bypasses across 48 vectors; the \texttt{NaN} amount denies at \texttt{money}; and authorized benign calls are allowed.

The companion measurement harness uses deterministic SHA-256 canary oracles, Wilson intervals, Newcombe paired confidence intervals, Benjamini-Hochberg FDR correction, temperature 0, sandboxed fixtures, and non-routable sinks. No live third-party service is attacked and no real payment, secret, or egress occurs.

\section{Conclusion}
\label{sec:conclusion}

Agentic frameworks commonly expose tools through capability gating and schemas. That is necessary, but insufficient. Once an LLM reads untrusted content, the model-emitted tool call is not an authorization decision. The audited default paths in LangChain/LangGraph, LlamaIndex, and the Stripe Agent Toolkit leave that decision to the integrator.

The fix is complete mediation at the tool-call boundary: a deterministic fail-closed PDP/PEP that checks each proposed action and argument value against out-of-band policy. \sg{} is one reference implementation of that boundary. It does not cure prompt injection. It makes the compromised model's reachable actions bounded by policy, reproducibly and before side effects.

\balance
\end{document}